\documentclass[12pt]{article}
\usepackage{graphicx} 
\usepackage{cite}

\def\nsprime{NS$^\prime$}
\newcommand{\abs}[1]{\left | #1 \right |} 

\begin{document}
\begin{titlepage}

\begin{flushright}
PUPT-1797,MIT-CTP-2749, \\
hep-th/9806092 \\
\end{flushright}

\begin{center}
\vskip3em
{\large\bf Brane Boxes: Bending and Beta Functions}

  \vskip3em
{L.\ Randall\footnote{E-mail:randall@mitlns.mit.edu},
\vskip .5em
{\it Center for Theoretical Physics\\Department of
Physics\\MIT, Cambridge, MA 02139 }\\
\vskip .5em
Y.\ Shirman\footnote{E-mail:yuri@feynman.princeton.edu}	 and\ R.\ von
Unge\footnote{E-mail:unge@feynman.princeton.edu}}\\ \vskip .5em 
{\it Department of Physics\\ Princeton University\\ Princeton, NJ
08544, USA\\}
	
\vskip2em
\end{center}

\vfill

\begin{abstract}
\noindent We study the type IIB brane box configurations recently
introduced by Hanany and Zaffaroni.  We show that even at finite
string coupling, one can construct smooth configurations of branes
with fairly arbitrary gauge and flavor structure.  Limiting our
attention to the better understood case where NS-branes do not
intersect over a four dimensional surface gives some restrictions on
the theories, but still permits many examples, both anomalous and
non-anomalous.  We give several explicit examples of such
configurations and discuss what constraints can be imposed on
brane-box theories from bending considerations.  We also discuss the
relation between brane bending and beta-functions for brane-box
configurations.

\end{abstract}

\vfill
\end{titlepage}

\section{Introduction}
Recently, following the ideas of \cite{HW}, it has become possible
to investigate non-per\-tur\-ba\-tive properties of supersymmetric
gauge theories by studying brane configurations in string and M theory
\cite{kutasov, wittensolution, ozetal, review}.  However,
constructing chiral $N=1$ supersymmetric gauge theories
has proven more difficult. Several classes
of chiral gauge theories were constructed in Refs. \cite{lpt,
lptmore, kutasovchiral, lowe, hananyetal}. An interesting 
development appeared in ref. \cite{hz}. In this paper it was
shown that in the limit of vanishing string coupling it is
possible to construct a large class of  chiral theories (including models
known to exhibit dynamical supersymmetry breaking). One
would like to understand the  dynamics of these brane
constructions at non-zero coupling. In this paper,
we take a step in this direction and discuss  
some of the properties of the brane configurations introduced in
\cite{hz}.

\begin{figure}[tc]
\centering
\includegraphics[height=2in]{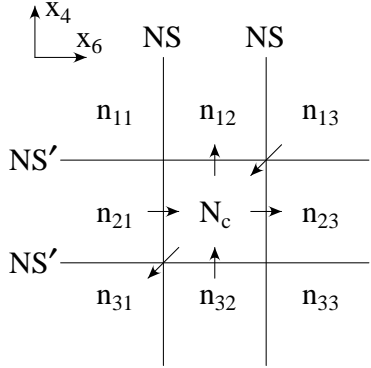}
\caption{A general brane-box configuration.
\medskip}
\label{fig:general} 
\end{figure}

Let us briefly recall the setup of the model. Following \cite{hz},
we consider field theory on D5-branes with world-volume in
dimensions $(012346)$. To obtain a four-dimensional low energy
effective theory, we need these D-branes to be finite in
two dimensions. Therefore we will let them end on NS5-branes with
world volume in $(012345)$ and \nsprime5-branes with
world-volume in $(012367)$. In addition to dimensional reduction
from 6 to 4 dimensions, the inclusion of NS-branes also reduces
the amount of  supersymmetry
 to $N=1$, and in general gives rise to   chiral theories.
The basic configuration with one
gauge group is shown in Figure \ref{fig:general}. If we put $N_c$
D5-branes   of finite extent  into the box we obtain an  $SU(N_c)$ gauge
theory, while product group theories arise from configurations with several
finite boxes. If we add (semi-infinite) D5-branes to the
surrounding boxes this will give rise to  matter in the
fundamental representation given by strings between
``flavor'' and ``color'' D-branes 
as shown in Figure \ref{fig:general}. 
By convention, incoming arrows correspond to anti-fundamental
fields, while outgoing arrows correspond to fundamental
fields (the same rules apply to the determination of charges
under global symmetries). It is also possible to introduce
more general matter content by adding D7-branes and
orientifold 7-planes to the configuration.

The original \cite{hz} construction was done for zero string coupling.
In this limit, one did not have to worry about the intersection of the
D5 and NS- or \nsprime-branes; all branes are orthogonal. It is much
more interesting if one can understand the bending of the branes
at nonzero coupling. This is important because it
  should somehow reflect the dynamics of the
field theories they represent. Here we should distinguish
between bending from perturbative and non-perturbative effects;
we will only address the former in this paper. A further
motivation for the investigation of brane-bending is that
it can potentially provide restrictions on the class of allowed
theories. This is important because it appears
  that arbitrary
configurations of the type described above 
(including those leading to
anomalous field theories) are  allowed.  

In the case of  weak coupling, it is possible to consider
the bending of the NS- (\nsprime-) branes
and check if there exist continuous 
non-intersecting solutions. A first step
in this direction was made in \cite{gg}. It was required
that asymptotically (for large values of $x_7$) the angle under
which the \nsprime-branes bend (and their $x_4$ coordinate) should be
independent of the $x_6$ coordinate on the brane (an analogous
conditions on the NS-brane bending did not lead to an  independent
constraint). The constraint required that:
\begin{equation}\label{eq:ggconstraint}
\begin{array}{c}
	n_{21}=n_{11}-n_{12}+N_c\\
	n_{23}=n_{13}-n_{12}+N_c\\
	n_{31}=n_{33}+n_{11}-n_{13}\\
	n_{32}=n_{33}+n_{12}-n_{13}
\end{array}
\end{equation}
While these conditions are sufficient to guarantee anomaly freedom,
they also allowed only theories with
$N_f \ge N_c$. In particular, the constraints imposed by Eq.
(\ref{eq:ggconstraint}) do not allow theories
to flow in parameter space to pure $N=1$ SYM; neither do they
allow one to construct SYM theory itself, which is rather
unsatisfactory.

In this paper we will show that the conditions of \cite{gg} are
too restrictive by demonstrating the existence of smooth
solutions to the brane bending problem. The existence
of such solutions is to be
expected given the new configurations proposed in
\cite{last} which avoid this restriction, though it was not
explicitly demonstrated there why the conditions of \cite{gg}
are unnecessary.  The only constraint we derive from bending
arises from restricting brane intersections. Unfortunately,
while our ``relaxed'' conditions will exclude some anomalous
models, in general anomalous theories are still
permitted. Therefore, it remains an open question to find
consistency conditions on the brane-box configurations (or
understand how anomalies can be cancelled due to the inflow
from the bulk theory in the spirit of
\cite{inflow}). Presumably this problem is solved by a
better understanding of the underlying string degrees of freedom of
the theory.

We will also show a connection between the running of the gauge
coupling constant and the bending of the branes. We will first argue
that in some simple configurations, the bending is logarithmic at
large distances from the brane intersections, reflecting 4-dimensional
physics. While we do not yet know the detailed connection between the
running of the coupling constant and the brane bending in general
cases, we will argue that the brane bending correctly reflects the
leading term in the beta function.

The paper is organized as follows: In Section \ref{bending} we will
find solutions for the brane bending in several illustrative cases and
show that there exist smooth solutions for the brane in general
configurations. In Section \ref{restriction} we will discuss what
restrictions might be imposed on brane-box configurations from the
requirement that the NS- (\nsprime-) branes do not intersect over a
4-dimensional surface. In Section \ref{sec:run} we will discuss how
the logarithmic running of the 4-dimensional gauge coupling is
reflected in the bending of the NS- (\nsprime-) branes in some simple
configurations. In Section \ref{sec:beta} we will argue that for
general configurations the beta function is encoded in the bending of
the branes. We summarize our results in Section \ref{sec:concl}.

\section{Bending Branes}
\label{bending}
Consider an \nsprime-brane in flat space. Its dynamics is governed by a
Dirac-Nambu-Goto type action. Its equation of
motion derived from the
action is
  \begin{equation}\label{DBI}
    \frac{1}{\sqrt{-\det\gamma}}
    \partial_{a}\left(\sqrt{-\det\gamma}
    \gamma^{ab}
    \partial_{b}X\right) = 0,
  \end{equation}
where $\gamma_{ab}$ is the induced metric
$\partial_{a}X^{M}\partial_{b}X_{M}$. The equation reduces to the flat
space Laplace equation whenever $\gamma_{ab}$ is constant and
proportional to the identity. If some other branes end on the NS-brane
they will act as sources for the Laplace equation and modify the right
hand side of (\ref{DBI}).

\begin{figure}
\centering
\includegraphics[height=1.5in]{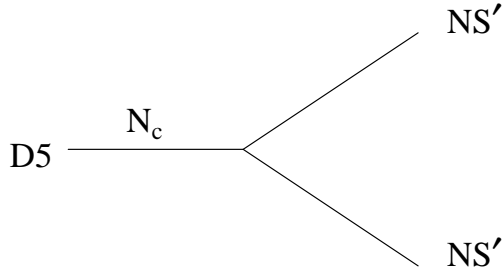}
\caption{D5-branes ending on a \nsprime 5-brane
\medskip}
\label{fig:fived}
\end{figure}

Now consider a D5-brane ending on an \nsprime-brane. As a result the
\nsprime-brane will buckle in the $x_4$ direction. Because of symmetry,
the $x_4$ coordinate of the \nsprime-brane depends only on $x_7$, the
distance from the D5-brane end point inside the
\nsprime-brane. Consequently, for large values of $x_7$ (far away from
the region where the source is strong), the buckling is determined by
the one dimensional flat space Laplace equation\footnote{One could
find a more precise solution by compactifying type IIB theory on a
circle and solving this configuration as M-theory on a torus
\cite{fivedcircle}.} \cite{ah}. The solution is given by:
   \begin{equation} \label{eq:linear} 
     x_4=k N_c \abs{x_7},
   \end{equation}
where $N_c$ is the number of D5-branes and $k$ is the proportionality
constant which vanishes in the limit $g_s \rightarrow 0$. (Note
that we have chosen integration constants differently from \cite{ah}
so that the configuration is more symmetric; see Figure
\ref{fig:fived}).

If the D5-branes also terminate on NS-branes, and thus are finite in
the $x_6$ direction, the situation changes. In that case the solution
of equation (\ref{DBI}) is no longer independent of $x_6$, and one
therefore has to solve a two dimensional Laplace equation.
The solutions for the branes will not surprisingly look very
different. For simplicity, we will solve the flat space
Laplace equation which is only an approximation to the true
equation of motion. As a consistency check, one can plug
the solution into the full equation of motion (\ref{DBI})
and verify that far away from the source we get back the
flat space Laplace equation. Generally, this means we trust
our solution better far from the intersection.

A relevant question to ask is what happens to supersymmetry when the
D5-brane is finite and the NS-brane therefore bends in two dimensions
as discussed in the previous paragraph. The answer can be found by
examining the $\kappa$-symmetry of the 5-brane \cite{martin}. We find
that supersymmetry is preserved when the following martix has zero
eigenvalues 
\begin{eqnarray} \label{eq:kappa}
 1 + \frac{\epsilon^{a_1\ldots a_6}}
          {\sqrt{-\det\left(\gamma + {\cal F}\right)}}\left(
  \frac{1}{3!} {\cal F}_{a_1 a_2}{\cal F}_{a_3 a_4} {\cal F}_{a_5 a_6}
 \right. \nonumber\\ \left.
 +\frac{1}{2!} {\cal F}_{a_1 a_2}{\cal F}_{a_3 a_4} \Gamma_{a_5 a_6}
 + {\cal F}_{a_1 a_2} \Gamma_{a_3\ldots a_6}
 + \Gamma_{a_1 \ldots a_6}\right).
\end{eqnarray}
Here $\Gamma_{a_1\ldots a_n}=\Gamma_{M_1\ldots
M_n}\partial_{a_1}X^{M_1}\ldots\partial_{a_n}X^{M_n}$, the pullback
of the space-time gamma matrices and ${\cal F}$ contains the field
strength of the world volume gauge field. When the 5-brane is flat
${\cal F}$ is zero and we get the usual equation for a supersymmetric
configuration. When the brane is curved we find that we have to turn
on some non-trivial ${\cal F}$ field to preserve supersymmetry in
analogy with \cite{CalMal}.

We now consider explicitly the form of the solution in
several relevant examples.

\begin{itemize}
\item  The first example we consider
is the one describing pure N=1 SYM theory.
We will solve for the bending of the \nsprime-brane.  In
our simplified discussion of the \nsprime-brane bending the only role
played by the NS-branes is to provide a point in the 6-7 plane on which the
D5-branes can terminate. To find the bending
of the \nsprime-brane caused by one D5-brane of length $2L$ ending on
it, we integrate the solution of the two-dimensional Laplace equation
for a point-like charge along the line $x_7=0$, $-L<x_6<L$ and find the
solution:
    \begin{equation} \label{eq:finitebox} 
    {x_4= 2 x_7 \left (\arctan ({L+x_6\over x_7})+ 
    \arctan ({L-x_6\over x_7}) \right) + \atop (L+x_6)
    \log((L+x_6)^2+x_7^2) + (L-x_6) \log((L-x_6)^2+x_7^2)}
    \end{equation} 
The first  conclusion to be drawn from Eq. (\ref{eq:finitebox})
is that there indeed exists a continuous and 
 smooth solution, and thus this brane configuration is
consistent. Note that when
viewed far from the intersection (large $x_7$), the solution
exhibits {\it logarithmic} (not linear) bending. This is to
be expected, since we no longer have an infinite line source
of charge. When viewed from far away, the finite line of charge
appears point-like; one expects logarithmic bending. This is
similar to the bending of the NS-branes in the type IIA
configuration \cite{wittensolution}. On the other hand, close to where the
D5-brane is located the solution looks almost linear.
To illustrate this point we plot a picture of a \nsprime-brane being
pulled by a D5-brane in Figure \ref{fig:finitebox}.
\begin{figure}[tc] 
\centering
\includegraphics{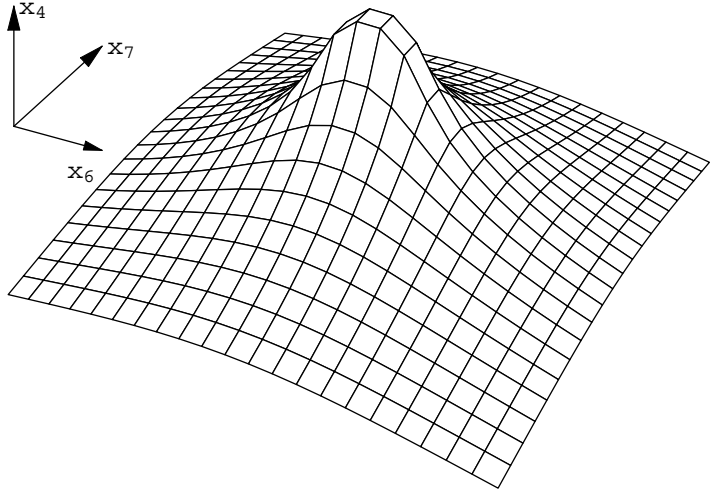}
\caption{The shape of \nsprime-brane pulled by a {\it finite}
D5-brane.
\medskip}
\label{fig:finitebox}
\end{figure} 

\item We next consider the solution for QCD with matter.
 We can add $N_L=N_R=N_f$ semi-infinite D5-branes on
the left and on the right (see Figure \ref{fig:nlnrmodel}). 
As was discussed
in \cite{hz} this corresponds to an $SU(N_c)$ gauge theory
with $N_f$ flavors in the fundamental representation. It is convenient
to find the solution for 3 finite boxes of length $L_L-L$, $2L$,
$L_R-L$, and then take a limit $L_L=L_R \rightarrow \infty $.
The bending of the \nsprime-branes is given by
    \begin{equation}
        \begin{array}{rl}
    x_4=&2(N_c-N_f) x_7 \left [\arctan ({L+x_6\over x_7}) + 
    \arctan ({L-x_6\over x_7})\right]+\cr
       &(N_c-N_f)(L-x_6) \log ((L-x_6)^2+x_7^2)+ \cr 
    &(N_c-N_f)(L+x_6)\log (L+x_6)^2+x_7^2 ))+ \cr
    &2 \pi N_f \abs{x_7},\cr
    \end{array}
    \end{equation}
where we subtracted constant terms.
At sufficiently large distances from the finite box, the bending is
approximately given by the superposition of the linear and logarithmic
terms. Again, the solution we find is
smooth. 
\begin{figure}[tc]
\centering
\includegraphics[height=1in]{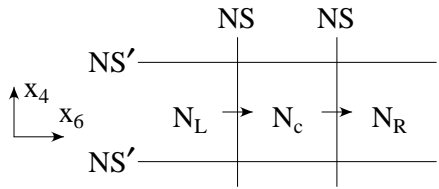}
\caption{A model with $N_L$ antiquarks and $N_R$ quarks
\medskip}
\label{fig:nlnrmodel}
\end{figure}

In the case we are considering, the
solution for the NS-brane bending is still
logarithmic at large distances. It is identical to the
solution in  Eq. (\ref{eq:finitebox}) with a factor of $N_c-N_f$
and an obvious relabelling of directions. 
When $N_f > N_c$, we find that the NS-branes
intersect. This is because the bending of one set of
NS-brane does not carry the full information about the breaking of
supersymmetry down to $N=1$. Thus the NS-brane intersection 
reflects the loss of asymptotic freedom
in the ``parent'' $N=2$ theory. We conclude that in addition
to SQCD with $N_f$ flavors the field
theory of this  brane configuration contains additional
degrees of freedom which have not been identified. We therefore
would not describe the $N_f > N_c$ theories in this
manner but instead would
construct configurations with $N_f > N_c$ by putting
some of the D5-branes in different boxes \cite{gg}.

\item The final configuration  we consider is a case with
no finite boxes, but two horizontal \nsprime-branes and a single vertical
NS-brane. One could similarly consider instead the example
with more vertical NS-branes which can be interpreted
as a gauge theory, but this simpler case is sufficient to illustrate
our point. Assume that the number of $D5$ branes on the left (right)
is $N_L~ (N_R)$. The solution contains infinite terms
of the form $x_6 (N_L\log L_L -N_R \log L_R)$. We can choose
$L_L$ and $L_R$ in such a way that these terms cancel, and
take the limit in a correlated fashion. The solution is then
\begin{equation}\label{eq:nlnenr}
\begin{array}{rl}
x_4= &x_6 \left(N_R \log((L-x_6)^2+x_7^2)- 
N_L \log((L+x_6)^2+x_7^2)\right)+\\
&2(N_R-N_L) x_7 \arctan({x_6\over x_7})+
(N_L+N_R) \abs{x_7}.
\end{array}
\end{equation}
Note that if either $N_L$ or $N_R$ were zero, we could not
cancel the divergence. In that case the \nsprime-brane would
become vertical in the 4-6 plane. This problem suggests that it
may be necessary for the D5-branes to end on D7-branes so that
all D5-branes in the configuration are finite. These are presumably
necessary if one is to describe correctly the moduli fields,
analogously to D6-branes in the type IIA theory \cite{review}.

	\begin{figure}[tc]
	\centering
	\includegraphics{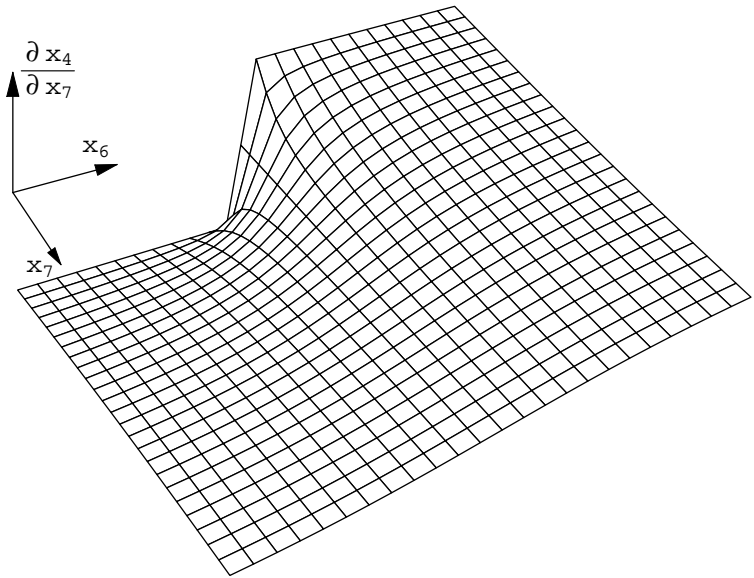}
	\caption{The {\it angle} of \nsprime-brane bending
	as a function of $x_6,x_7$
	\medskip}
	\label{fig:angle}
	\end{figure}

In \cite{gg}, it was argued that for $N_L \ne N_R$, there
would be a sharp transition as one moves along $x_6$, 
which would literally tear the branes apart at
large $x_7$ distances. As we can see from the solution in
Eq. (\ref{eq:nlnenr}), the situation is quite different.
We find that the transition region is smooth
everywhere and in fact becomes more smooth as one goes to large
$x_7$. This is illustrated in Figure \ref{fig:angle}
 where we have plotted the angle
at which the brane bends ($\frac{\partial x_4}{\partial x_7}$) as a
function of $x_6$ and $x_7$. As can be seen, there is a transition
between two different bending angles but it is smooth everywhere and
the branes do not break.

\end{itemize}

{}From these examples, the general behavior of the solution should be
clear. A general solution is a superposition of elementary solutions
and in our approximation NS- and \nsprime-branes bend independently.
For instance, if we consider many finite boxes in a row 
(corresponding to product group theories) the solution is given by a
linear superposition of many solutions from the first example;
again the general tendency should be clear. There is always
a smooth solution to the two dimensional Laplace equation which 
approximates the true answer. However, we have not yet considered
restrictions which might be imposed when there is more than one
NS- or \nsprime-brane which is bending, so that there can be
intersections which lead to new field theory degrees of freedom. We do
this in the next section.

\section{Restrictions from Brane Intersections and Their Solutions}
\label{restriction}
\begin{figure}
\centering
\includegraphics[height=2in]{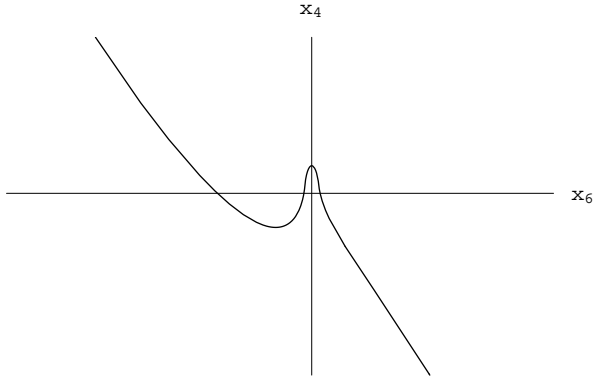}
\caption{A section of \nsprime-brane at $x_7=0$ in a model 
with anomalous matter content.
\medskip}
\label{fig:anomalous}
\end{figure}

In the previous section, we found that there exist smooth
solutions for the \nsprime- (NS-) brane shapes for arbitrary D-brane
configurations. Thus it appears that one can relax the conditions
imposed in \cite{gg}. Can one impose any restrictions on allowed
brane configurations by considering their bending? 
We will require that all brane configurations
are such that the NS- (\nsprime-) branes do not intersect over a
4-dimensional manifold. If the branes would intersect, there would be
additional light degrees of freedom living on the intersection and the
content of the effective field theory would be different.  Let us
illustrate the consequences of this requirement on an anomalous model
with $N_L\ne N_R$. Combining the solution for pure SYM theory in
Eq. (\ref{eq:finitebox}) with the solution for semi-infinite D5-branes in
Eq. (\ref{eq:nlnenr})   
we plot the shape of one \nsprime-brane at
$x_7=0$ in Figure \ref{fig:anomalous}. Since the other
\nsprime-brane bends in the opposite direction 
we find that two (classically) parallel
\nsprime-branes intersect over a 4-dimensional surface.
Thus the anomalous theory is problematic
as a brane configuration. We therefore we impose the requirement
$N_L=N_R$. Notice that this condition was imposed by hand from a
knowledge of the field theory in \cite{hz} but here we derive it from
the brane configuration itself. For any theory constructed out
of two NS-branes and two \nsprime-branes the above condition becomes:
\begin{equation}
\label{eq:newcondition}
\begin{array}{c}
n_{31}=n_{11}+n_{33}-n_{13},\\
n_{32}=n_{33}+n_{12}-n_{13},\\
n_{21}=n_{11}+n_{23}-n_{13}.\\
\end{array}
\end{equation}

It is easy to further generalize our conditions to an arbitrary
array of brane boxes constructed out of H \nsprime-branes
and V NS-branes. Requiring that branes do not intersect
imposes $H+V-1$ independent conditions on $(H+1)(V+1)$
numbers of D5-branes, which is far less than the $(H-1)(V-1)$ 
conditions imposed in \cite{gg}. Yet we note that these conditions
exclude some of the interesting models, including the $3-2$ model of
dynamical supersymmetry breaking.

It is, however, not quite clear to us whether even these less
restrictive conditions imposed above are necessary. For example,
instead of introducing matter by having semi-infinite D5-branes one
could allow them to end on D7-branes as noted earlier. Then moving the
D7-branes sufficiently close to the gauge theory box, one could
guarantee that there are no 4-dimensional intersections of NS-
(\nsprime-) branes. If this were possible, then the physics would
depend on the $x_6$ position of the D7-branes and the $x_4$ position
of the D7$^\prime$ branes. This is in contrast to the role played by
the D6-branes in type IIA configurations, where holomorphic quantities
do not depend on the $x_6$ coordinate of the D6-branes. This might be
a reasonable difference, since it is difficult to identify the
holomorphic structure of the field theory in the brane box
configuration. We should mention however that the D7-branes change
the geometry of space, introducing a deficit angle, so our flat space
arguments may not be valid in their presence.

\begin{figure}[tc]
\centering
\includegraphics[height=2in]{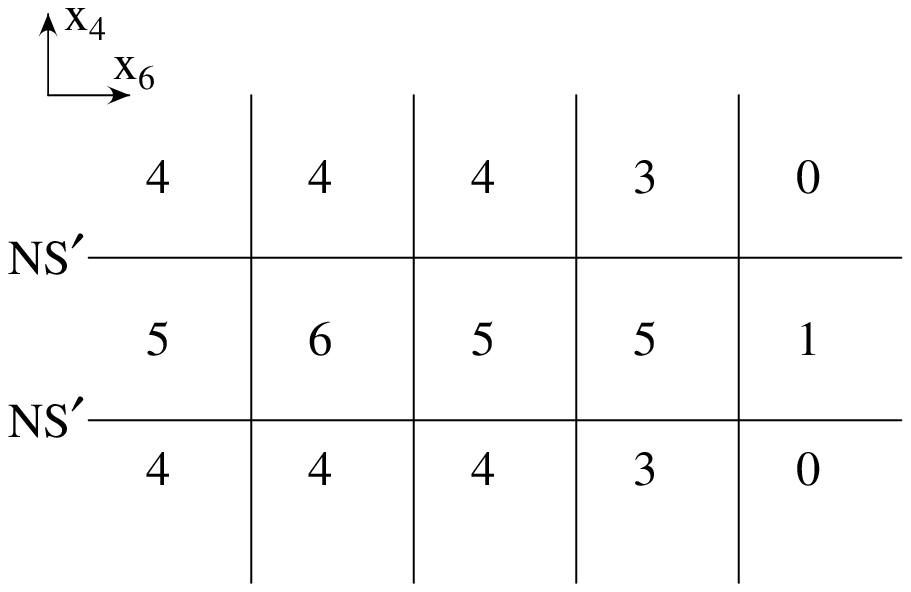}
\caption{A non-intersecting anomaly free configuration which does
not satisfy conditions of Eq. (\ref{eq:ggconstraint}).
\medskip}
\label{fig:notgg}
\end{figure}

While we lack a complete understanding of the conditions which should
be imposed on brane configurations due to bending, we nonetheless can
go on and consider implications of the above conditions together with
anomaly freedom conditions imposed {\it by hand}. The total number of
conditions would then be $(H-1)(V-1)$ which is the same as the number
of conditions in \cite{gg}. It turns out that the combination of our
constraints and anomaly cancellation is always satisfied by the
conditions of \cite{gg}. However, in certain cases some of the anomaly
freedom conditions are redundant, allowing theories with
non-intersecting branes which do not satisfy conditions \cite{gg}. We
have seen an example of this in supersymmetric QCD.  In Figure
\ref{fig:notgg}, we give another example of an anomaly free, consistent
brane configuration which does not satisfy the condition \cite{gg}.

Finally, there is an additional subtlety. Suppose we had
started with a configuration in which the $i$'th and $(i+1)$'th
NS (or \nsprime) branes
remain parallel (even if they bend), 
so that the  distance between them does not depend on the
coordinate in the 4-5 (6-7) plane. 
Then we can add $\Delta N_{i-1}$, $\Delta
N_i$, and $\Delta N_{i+1}$ D5-branes to the  corresponding (finite) boxes.
As a result, the distance between the  NS- (\nsprime-) branes will change 
logarithmically at large values of $x_4,x_5$ ($x_6,x_7$).
If $\delta N_{i-1}+\delta N_{i+1} > 2 \delta N_i$, the two NS-
(\nsprime-) branes under
consideration will intersect. We therefore exclude such
configurations.

\section{The Running Coupling Constant}
\label{sec:run}

Our discussion of brane bending is also useful for 
understanding the relation of the running coupling
constant to the brane configuration. Due to 
dimensional reduction, the 4-dimensional
coupling constant is related to the area of the D5-brane
bounded by the NS-branes. Usually, the running of the coupling
constant is associated 
with the bending of the NS-branes. However, since the
intersection between the D5-brane and the NS- (\nsprime-)
brane is 4-dimensional, naively one would expect that
the bending  is linear, which would not reflect
the running of the coupling constant in 4 dimensions.

As we have seen, in a pure SYM configuration and at sufficiently large
distances from the finite D5-branes, the solution (and therefore, the
distance between two NS-branes) changes logarithmically:
\begin{equation} 
    \Delta x_4 \propto 2N_c \log R, 
\end{equation}
where $R= \sqrt{x_6^2+x_7^2}$.  It is tempting to identify
this
logarithmic bending with the properties of a 4-dimensional
theory in analogy with
\cite{wittensolution}. However,
the coefficient of the log does not reproduce the beta function
coefficient correctly.  In fact the coefficient gives the correct value
for $N=2$ SYM. This is not surprising because our configuration has
originated as an $N=2$ theory, and has been reduced to $N=1$ due to
the presence of NS-branes.  This is similar to the situation in the
type IIA constructions \cite{kutasov} of $N=1$ SYM. These models were
constructed by rotating one of the NS-branes in the $N=2$ theory. The
bending of each of the NS-branes is determined only by the number of
D4-branes attached to it, and reflects the $N=2$ running coupling
constant \cite{ozetal, kaplunovsky}. One infers the beta function
coefficient and the running of the coupling constant of the $N=1$
theory by taking into account the mass of the adjoint chiral multiplet
which is not reflected in the bending of the branes.  Similarly in our
case, the running coupling constant should be given by some function of
the distances between both NS and \nsprime-branes.

\begin{figure}
\centering
\includegraphics[height=2in]{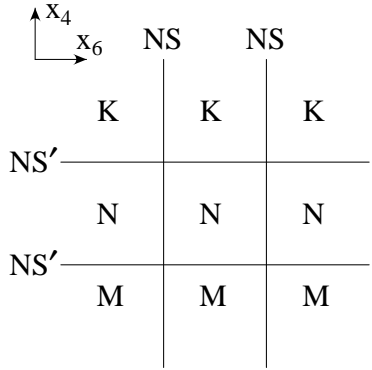}
\caption{A configuration corresponding to  an $SU(N)$ gauge group with
$N+M+K$ flavors in which the NS-branes do not bend.
\medskip}
\label{fig:bzero}
\end{figure}
Nevertheless,  there do exist
configurations with a simple relation
between brane bending and the running gauge coupling. 
As has been argued in \cite{brandhuber}, this is the case
in the configurations where one set of NS-branes
does not bend. In \cite{brandhuber} the
one-loop coefficient $b_0$ of the beta function for such models
(see Figure \ref{fig:bzero}) was determined. 
The model in Figure \ref{fig:bzero} corresponds to an
$SU(N)$ gauge theory with $N_f=N+M+K$ flavors.
Let us give an argument for the running coupling constant which
is somewhat different  from that in \cite{brandhuber}. 
The gauge coupling in the model is given by ${1\over g^2}=
{\Delta x_4 \Delta x_6 \over g_s l_s^2}$. The distance
between the  \nsprime-branes depends linearly on $x_7$: 
\begin{equation}\label{eq:distance}
\Delta x_4 (x_7)= \Delta x_4 + (3N-N_f) x_7.
\end{equation}

Let us reconnect $n$ D5-branes in the middle row to
form an infinite D5-brane and move them away in $x_7$, 
thus Higgsing
the theory to $N-n$ colors and $N_f-n$ flavors.
If we require that the  asymptotic positions 
of the  NS- and \nsprime-
branes do not change (this would require an infinite amount
of energy from a 4-dimensional point of view), then the 
area bounded by the  NS- and
\nsprime-branes (and, therefore  the gauge coupling) will change:
\begin{equation}
  \Delta A \propto 2n x_7
\end{equation}
where $x_7$ is the final position of the  D5-branes. The coefficient
in front of $x_7$ correctly reproduces the {\it change} in the
one-loop beta-function coefficient $b_0$. We could give analogous
arguments for turning on mass terms by reconnecting an arbitrary
number of D5-branes in the lower and/or bottom rows. Thus it is
tempting to identify the coefficient in front of $x_7$ in
Eq. (\ref{eq:distance}) with $b_0$. However, this would imply
linear running of the gauge coupling, which seems to reflect
that the origin of the theory is a five-dimensional theory. 
In fact, we can Higgs
the theory to take the D-branes off the NS-branes so that the theory
really seems to be five-dimensional\footnote{Note that in the anomalous theory 
discussed in Section \ref{restriction} 
some of the semi-infinite D5-branes can not be reconnected into infinite ones,
and the theory can never look truly five-dimensional. This is in an agreement 
with the fact that one does not expect inconsistencies in a five-dimensional 
configuration.}.

As we have noted before, even some potentially  
problematic configurations may
be acceptable (as far as the bending goes) 
if all the  D5-branes are finite (which requires that
flavor D5-branes end on D7 or D7$^\prime$ branes). 
In the configuration of Figure \ref{fig:bzero}, we therefore introduce
D7-branes. Then asymptotically, the  distance between the 
\nsprime-branes changes {\it logarithmically} with the coefficient
appropriate for $N=1$ theory:
\begin{equation}
\Delta x_4 \propto (2N-M-K) \log R = (3N_c-N_f) \log R.
\end{equation}
The arguments we have given above for matching couplings
of the  high energy and low energy theories are still valid, and
thus we see the correct logarithmic running of the coupling constant.

\section{The Beta Function}
\label{sec:beta}
While we cannot extract the running coupling constant from brane
bending in an arbitrary configuration, it is possible to constrain the
form of the beta function (or more precisely, the one-loop coefficient
of the beta function $b_0$) in the brane-box theories. The argument
goes as follows. Assume that the beta function is an arbitrary
function of $N_{c}$ and $N_{f}$ ($b_{0} = f(N_c , N_f )$) and that it
can be related to the bending of the branes.  A precise knowledge of
this relation is not necessary for the argument. From the assumption
we conclude that any deformation of the brane configuration that does
not change the bending of the branes cannot change the beta
function. In particular we can add one D5-brane that covers all the
boxes. Doing so we see that $N_c \rightarrow N_c + 1$ and that $N_f
\rightarrow N_f +3$. The requirement that the beta function is
invariant under this tells us that it is a function of the particular
combination $3N_c - N_f$ which is indeed correct for $SU(N_c)$ gauge
theories with $N_f$ chiral multiplets in the fundamental
representation.

Note that the same type of argument could be applied to the type IIA
configurations studied in \cite{wittensolution}. There one realizes
$N=2$ supersymmetric gauge theories on the world volume of D4-branes
suspended between NS-branes. If one adds a D4-brane that extends from minus 
infinity to plus infinity in $x^{6}$, one does not change the bending
of the NS-branes but one does indeed change the number of colors by
one and the number of flavors by two giving the invariant combination
$2N_c-N_f$ which agrees with the $N=2$ beta function.

\smallskip
So far, using the assumption that the brane bending encodes the beta
function,  we have established that (the one loop coefficient of) the
beta function is an arbitrary function of $3N_c-N_f$: 
$b_0=f(3N_c-N_f)$. Assuming that this function is universal, we can
determine its exact form by explicitly calculating its value in some
simple cases. In Section \ref{sec:run} we were able to directly relate
the change in brane bending of the configuration given in Figure
\ref{fig:bzero} to the change of the value of $b_0$. This tells us
that $b_0 = 3N_c - N_f + c$ where $c$ is an undetermined
constant. Finally we note that configurations without brane bending
corresponds to field theories where $b_0=0$ \cite{finite}. Such
configurations have $N_f = 3N_c$ and thus we find that $c=0$.

In the presence of orientifolds, the identification of the degrees of
freedom is less well understood. However, we believe that the argument
given above should still apply in this case. The argument could in
fact be used as a consistency check on any proposed counting of
degrees of freedom in the presence of an orientifold plane.

\section{Conclusions}
\label{sec:concl}
In this paper, using the
fact  that the bending of  NS-branes in brane-box
theories is   determined by a two dimensional Laplace
equation, we have demonstrated the existence
of consistent solutions. A D5-brane ending on the NS-brane looks like a line of
charge and a line of charge of finite extent looks like a point charge
at long distance and thus gives rise to logarithmic bending rather
than linear bending. This means that we can relax some of the
consistency conditions on these theories proposed in \cite{gg} since
they were derived assuming that all branes bend linearly. At long
distance the brane bends as a smooth solution of the two dimensional
Laplace equation and the branes do not break. What can happen though
is that two of the NS-branes intersect at some distance. In that case
there can be new massless states that has not been taken into account
in the field theory and it does not capture the
physics on the branes correctly.

The requirement that branes do not intersect is much less restrictive
than the condition given in \cite{gg}. In particular, we find
configurations of branes that do not intersect but that give rise to
field theories that are anomalous. This shows that there is no direct
relation between field theory anomalies and the brane
bending. Presumably to understand the anomalies better one would have
to look at non-conservation of 
Ramond-Ramond charge in the brane set-up.

We also argued that if all ``flavor'' D5-branes are finite ending on
D7 and/or D7$^\prime$ branes, and if the D7- (D7$^\prime$-) branes are
sufficiently close, there may be even less restrictions. We have
also seen that in some cases, the D7-branes may be required in order
to have a truly 4-dimensional theory. However, we have observed a
mysterious dependence on the $x_6$ ($x_4$) coordinate of the D7-
(D7$^\prime$-) branes.  Also, the D7-branes introduce a deficit angle
into the geometry and there might be restrictions on the allowed
number of D7-branes. The physics associated with the introduction of
D7-branes into the system requires further investigation.

We also showed that in some simple configurations and in the presence
of D7 branes, the logarithmic running of the gauge couplings can be
simply read off from the brane bending. Although it is not manifest
how this relation works in general, we have demonstrated that the
bending of the branes does correctly reflect the beta function
dependence on $N_c$ and $N_f$. We commented that this should also be
true for theories with orientifolds. The argument makes it clear that
the bending of the branes does encode the beta function and provides a
nice consistency check on the brane-box theories since it relies only
on the way flavors are counted in these theories. It would be
interesting to identify more precisely how the branes encode this
information about the beta function. Also of importance ultimately is
to see how the branes incorporate quantum effects. Since these are
presumably most important in the vicinity of the intersection of D5
and NS- or \nsprime-branes, one might require a more exact solution to
the equations of motion in order to explore non-perturbative gauge
dynamics with brane-box theories.

\section{Acknowledgments}
We are grateful to Martin Gremm, Michael Gutperle, Amihay Hanany, Anton
Kapustin, Emanuel Katz, Juan Maldacena, and Alberto
Zaffaroni for useful discussions. Y.~S. and R.~v.~U. would like to
thank MIT for hospitality during the completion of this paper.
The work of Y.S. was supported in part by an NSF grant PHY-9157482
and a James S. McDonnel Foundation grant No.91-48. The work of
L~.R. is supported in part by DOE under cooperative agreement
\#DE-FC02-94ER40818, NSF Young Investigator Award, Alfred P. Sloan
Foundation Fellowship.

\newcommand{\NPB}[1]{{\sl Nucl. Phys.} {\bf B#1}}
\newcommand{\PLB}[1]{{\sl Phys. Lett.} {\bf B#1}}
\newcommand{\PRL}[1]{{\sl Phys. Rev. Lett.} {\bf B#1}}
\newcommand{\PRD}[1]{{\sl Phys. Rev.} {\bf D#1}}
\newcommand{\CQG}[1]{{\sl Class. Quant. Grav.} {\bf #1}}
\newcommand{\ATMP}[1]{{\sl Adv. Theor. Math. Phys.} {\bf #1}}
\newcommand{\JHEP}[1]{{\sl JHEP} {\bf #1}}

\end{document}